\documentstyle[aps,twocolumn,prb]{revtex}
\begin{document}
\draft

\title{
Columnar dimer and plaquette resonating valence bond states\\
of the quantum dimer model
}

\author{P. W. Leung\cite{email} and K. C. Chiu}
\address{Department of Physics,
Hong Kong University of Science and Technology\\
Clear Water Bay,
Hong Kong
}
\author{Karl J. Runge}
\address{Department of Physics,
Lawrence Livermore National Laboratory, University of California\\
Livermore, California 94551
}
\date{\today}
\maketitle
\begin{abstract}
We study the nature of the ground state
of the quantum dimer model proposed by
Rokhsar and Kivelson by diagonalizing the Hamiltonian of the model
on square lattices of size $L\times L$, where $L\leq 8$, with periodic
boundary conditions.  Finite-size scaling studies of the columnar
order parameter and the low lying excitation spectrum
show no evidence of a dimer liquid
state in any finite region of the zero temperature phase diagram.
In addition, we find evidence of a transition from the columnar
dimer state to an intermediate state at a negative value
of $V/J$.  This state is identified to be the
plaquette resonating valence bond (RVB) state.  The energy gap of this state 
vanishes as a power law of $L$.
It exhibits columnar dimer order,
but has disorder {\it within} the dimer columns.  This state persists up
to $V/J<1$, and the system changes to a dimer liquid state only at $V/J=1$.
\end{abstract}

\section{Introduction}

The quantum dimer model (QDM)
was first introduced by Rokhsar and Kivelson (RK)\cite{rk88}.  
It was proposed as an alternative description of the
non-N\'eel state of the spin-$\frac{1}{2}$ Heisenberg antiferromagnet
on a square lattice.
In a state with exponentially decaying spin-spin correlation
and a large energy gap separating the spin excitations from the
singlet ground state, RK argued that the low energy physics
is contained in the ``short-range resonating valence bond (RVB)''
states, spanned by the set of nearest-neighbor valence bond states.
Nearest-neighbor valence bonds are called dimers.  The basis
set thus consists of all possible dimer configurations
at closest-packing (all sites form exactly one dimer with one of their
nearest neighbors).  RK also argued that the non-orthogonality
of these basis states can be absorbed by defining a 
phenomenological Hamiltonian with short range dimer interactions.
The relevance of this quantum dimer model to the frustrated
quantum antiferromagnet was supported by other independent
studies.  Large-$N$ expansion,\cite{meanfield}
series expansion,\cite{series} and numerical diagonalization\cite{j1j3}
have shown that a spin-Peierls
state exists in some region of the phase diagram where the N\'eel
state is unstable.  This spin-Peierls state corresponds to the
state in the QDM where the dimers are frozen into
a columnar pattern (Fig.~\ref{dimer}(a)). 
Recently the QDM has been generalized to the kagom\'e
lattice by Zeng and Elser.\cite{ze95}

The Hamiltonian of the QDM proposed by RK is
\begin{eqnarray}
{\cal H}=\sum_{\rm plaquettes}\biggl[&-&J\biggl(|\parallel\,\rangle
\langle=|+{\rm H.c.}\biggr)\nonumber \\
&+&V\biggl(|=\rangle
\langle=|\ +\ |\parallel\,\rangle\langle\,\parallel| \biggr)\biggr].
\label{hamiltonian}
\end{eqnarray}
$\parallel$ and $=$ represent parallel dimers on the opposite sides of a
square plaquette.  The first term of ${\cal H}$ is the dimer kinetic energy
operator while the second term is the potential energy.
The space spanned by all close-packed dimer configurations
on a square lattice can be divided into distinct topological
sections characterized by a pair of conserved
winding numbers $(\Omega_x,\Omega_y)$ (refer to Appendix~\ref{app:wind}).
On an $L\times L$ square lattice, the allowed winding numbers
are $-L/2\le \Omega_x,\Omega_y\le L/2$.
At $V/J>1$, the exact ground state of the QDM is the staggered
dimer solid shown in Fig.~\ref{dimer}(b).  It  is four-fold degenerate
and has winding numbers $(\pm L/2,0)$ and $(0,\pm L/2)$.
At $V/J \ll -1$, the ground state develops columnar order as shown in
Fig.~\ref{dimer}(a).
At $V/J=1$, the lowest energy states in all topological
sectors are zero-energy eigenstates of ${\cal H}$.
They are equal superpositions of all the dimer configurations
in their sectors.  Therefore, the ground state at $V/J=1$
is a dimer liquid, and any ground state dimer correlation function
can be calculated exactly by the method of Fisher and
Stephenson\cite{fs63} for the classical dimer problem.
Consequently this state is called the FS state.
The ground state properties of the QDM in the range
$0\le V/J<1$ are  not so clear.
There are three main questions: (a) is the ground state in this range ordered
(dimer solid) or disordered (dimer liquid)? 
(b) If it is ordered, what is the nature of this order?
And (c)
if it is disordered,
is there a gap in the low lying excitation spectrum?
RK suggested that a dimer liquid state can exist 
in a finite region of $V/J<1$.
By numerically diagonalizing the QDM on square lattices up to
$6\times6$, Sachdev\cite{s89} found evidence for the columnar
state, but no evidence
for a dimer liquid state
at $V/J\not= 1$.  However, due to finite-size effects,
he did not rule out the possibility
that dimer liquid exists over a finite range $\kappa<V/J\le1$, and estimated
a lower bound $0.5<\kappa$.
On the other hand, some analytic studies have suggested
the existence of a dimer liquid state at $V/J\ne 1$.
By mapping the QDM to a roughening problem,
Levitov\cite{l90} showed that a dimer liquid with gapless
excitations may exist.
Orland\cite{o94} formulated the QDM as a system of non-interacting
fermionic strings, and showed that at $V/J=0$,
the ground state is a dimer liquid.  From this he inferred
that the ground state is a dimer liquid for $0\le V/J\le1$,
with a phase transition to a dimer solid at some $V/J<0$.
We also note that another kind of dimer order has been
proposed,\cite{zu96} although in a different context.
This is a plaquette RVB state
as shown in Fig.~\ref{dimer}(c), where every other
plaquette  is in the
$|\parallel\,\rangle$ or $|=\rangle$ state with equal
probability and independent of other plaquettes.
In this paper, we extend Sachdev's calculations to 
an $8\times8$ square lattice.  We aim to identify the
different phases of the QDM in the range $V/J<1$.

\section{Numerical calculations}
\label{sec:numerical}

Using the Lanczos algorithm, we diagonalize the Hamiltonian ${\cal H}$
in the range $-1<V/J<1$
on an $L\times L$ square lattice with periodic boundary conditions,
where $L=4$, 6, and 8.  The basis states are all possible dimer
configurations at closest packing.  The number of basis states for
a given $L$ can be evaluated\cite{fs63} analytically.
Enumerating all the
dimer configurations is much more  difficult.
Nevertheless, we have devised an efficient algorithm which
enumerates all the 300 million dimer configurations
on the $L=8$ lattice within four hours using an HP 735 workstation
(see Appendix~\ref{app:count}).
Using translational symmetry, we can reduce the number of basis
states on an  $L\times L$ lattice by a factor of about $L^2$.
For $V/J<1$, the ground state has momentum ${\bf k}=(0,0)$, and
winding numbers $(0,0)$.  By restricting the basis states
to the topological sector with winding numbers $(0,0)$, the number of basis
states can be further reduced by a factor of about 2.
To calculate the
ground state eigenvector, we restrict ourselves to the subspace
with momentum ${\bf k}=(0,0)$ and winding numbers  $(0,0)$.
The number of basis states in this subspace is about $2.4$ million.
This makes the numerical diagonalization possible on a workstation.

\section{Symmetries of low lying excited states}
\label{sec:energy}

To study the different phases in the range $-1<V/J<1$,
we first study the symmetries of the low lying excited
states.
They are important in finite-size studies.
If a system possesses a broken symmetry
in the thermodynamic limit, the ground state of the finite
system will still be totally symmetric.  In this case the
ground state expectation of the appropriate
order parameter will have long-range correlations, 
and there will exist low lying excited states
with the appropriate symmetries.
The columnar dimer state shown in Fig.~\ref{dimer}(a) 
has
winding numbers $(0,0)$ and is
four-fold degenerate.  These degenerate states can 
be combined to form four states,
two with momentum $(0,0)$, and two with momenta $(\pi,0)$
and $(0,\pi)$.
Consequently in a finite system which possesses columnar
order in the thermodynamic limit, the state with momentum $(0,0)$
will be the ground state and the others will appear as low lying excited
states, which are degenerate with the ground state in the
thermodynamic limit.
Note that the same is true for the plaquette RVB state
shown in Fig.~\ref{dimer}(c).
Therefore, low lying excited states with momenta $(\pi,0)$
and $(0,\pi)$ may indicate that columnar or plaquette
RVB order exists in the ground state.

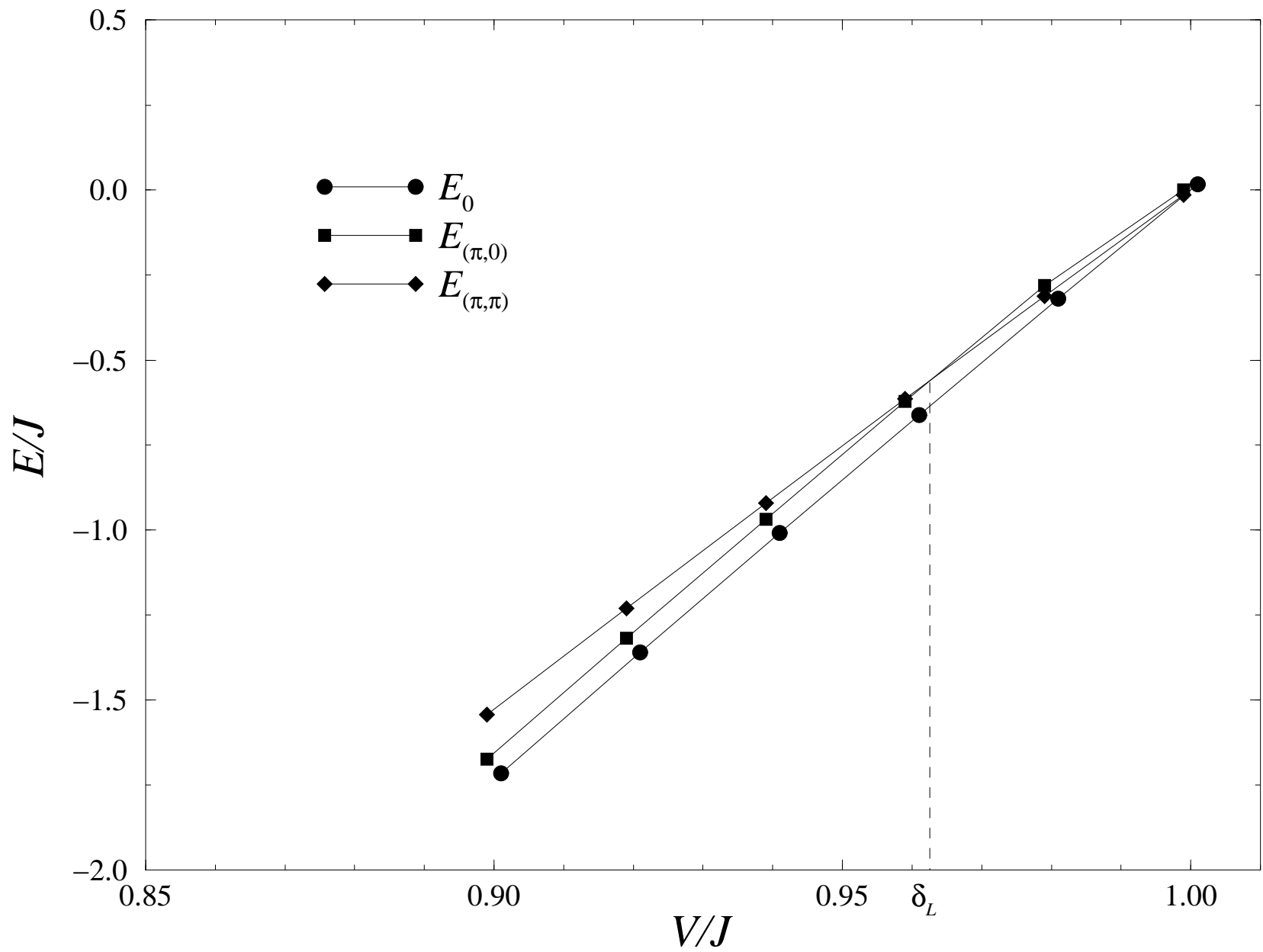
\begin{figure}
\caption{Energy levels $E_0$, $E_{(\pi,0)}$, and $E_{(\pi,\pi)}$
for the $L=8$ system.  All other energy levels are higher than these
three and are not shown.}
\label{level}
\end{figure}

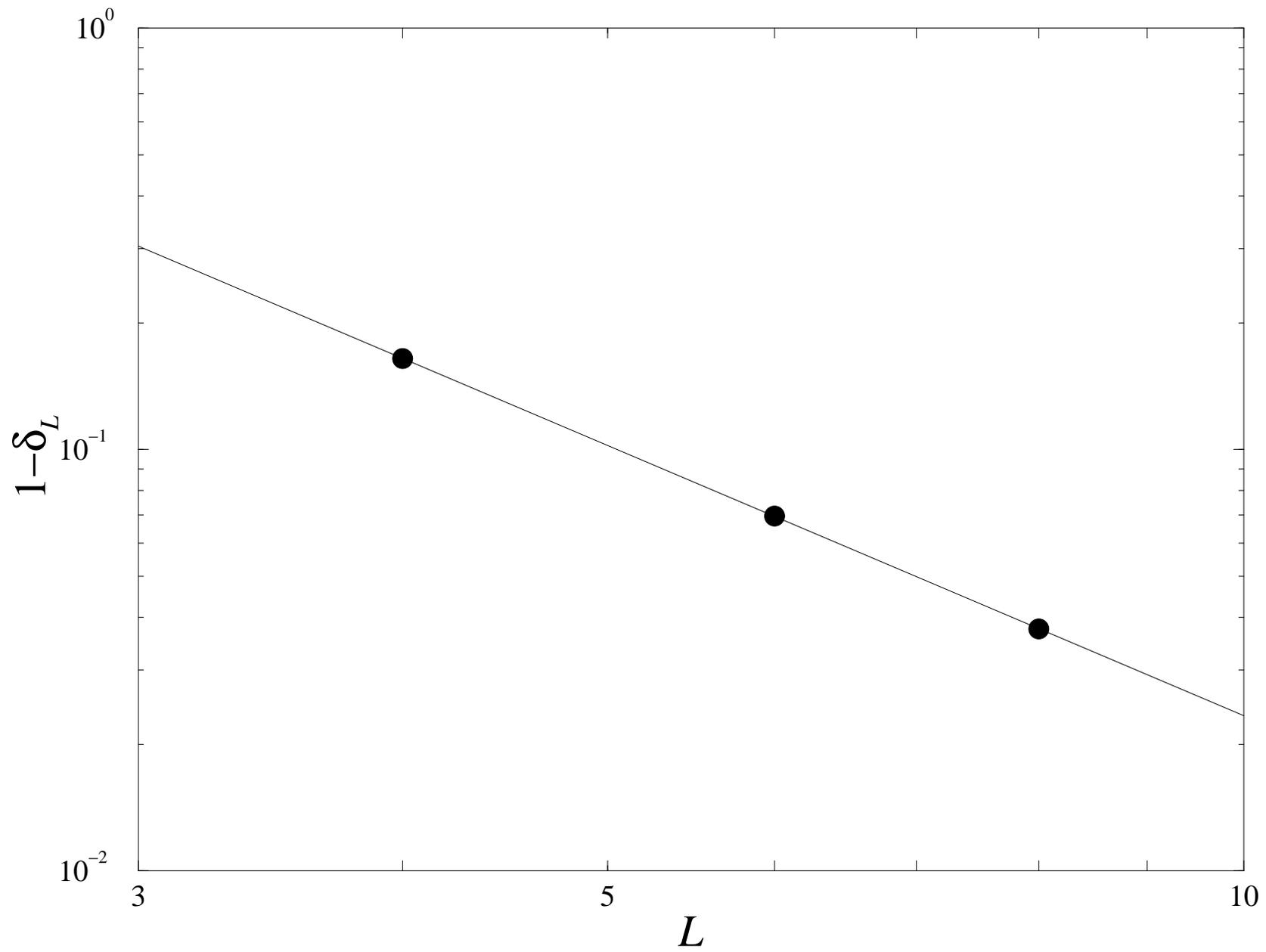
\begin{figure}
\caption{$(1-\delta_L)$ vs $L$.  The straight line is the best fit to
the data, $1-\delta_L=3.1544\,L^{-2.13}$.}
\label{deltaL}
\end{figure}

We calculate the energies of the lowest few eigenstates of
the QDM on the $L=4$, 6 and 8 lattices for $-1\le V/J<1$.
All three lattices show qualitatively the same picture.
The ground state $E_0$ always has zero momentum and winding numbers,
as mentioned in section~\ref{sec:numerical}.
Except when $V/J$ is close to 1, the first excited state is
degenerate with momenta $(\pi,0)$
and $(0,\pi)$, and has zero winding numbers.
We call this state $E_{(\pi,0)}$.
In the same region, the next excited state is degenerate
with momenta $(0,0)$ and $(\pi,\pi)$, and has
winding numbers $(\pm1,0)$ and $(0,\pm1)$.
We call this state $E_{(\pi,\pi)}$.
In Fig.~\ref{level} we plot the energies of the states
$E_0$, $E_{(\pi,0)}$ and $E_{(\pi,\pi)}$ in a small range
of $V/J$ close to 1 for the $L=8$ system.
We can see that there exist a $\delta_L$ (which depends on the size $L$)
close to but less than 1, such that
at $V/J=\delta_L$, $E_{(\pi,\pi)}$ and $E_{(\pi,0)}$ cross each other
and $E_{(\pi,\pi)}$ becomes the first excited state for $\delta_L<V/J<1$.
This crossing of energy levels may imply the existence of a new
ground state in the region $\delta_L<V/J<1$.  
However, finite-size
extrapolation analysis on $\delta_L$ shows that this is not
the case.
As shown in Fig.~\ref{deltaL}, $\delta_L$ approaches 1 as a power law
of $L$,\cite{powerlaw}
\begin{equation}
(1-\delta_L)\propto L^{-2.13}.
\end{equation}
In other words, such a spurious
energy level crossing is only a finite-size effect.
It does not
occur in the thermodynamic limit and no transition is likely
to occur at $V/J$ close to 1.
Therefore, we expect that the $E_{(\pi,0)}$ state is always the lowest
excited state in the range $-1\le V/J<1$.  This is consistent with,
but does {\it not} imply, the existence of long-range
order (LRO) in the QDM at $-1\le V/J<1$.

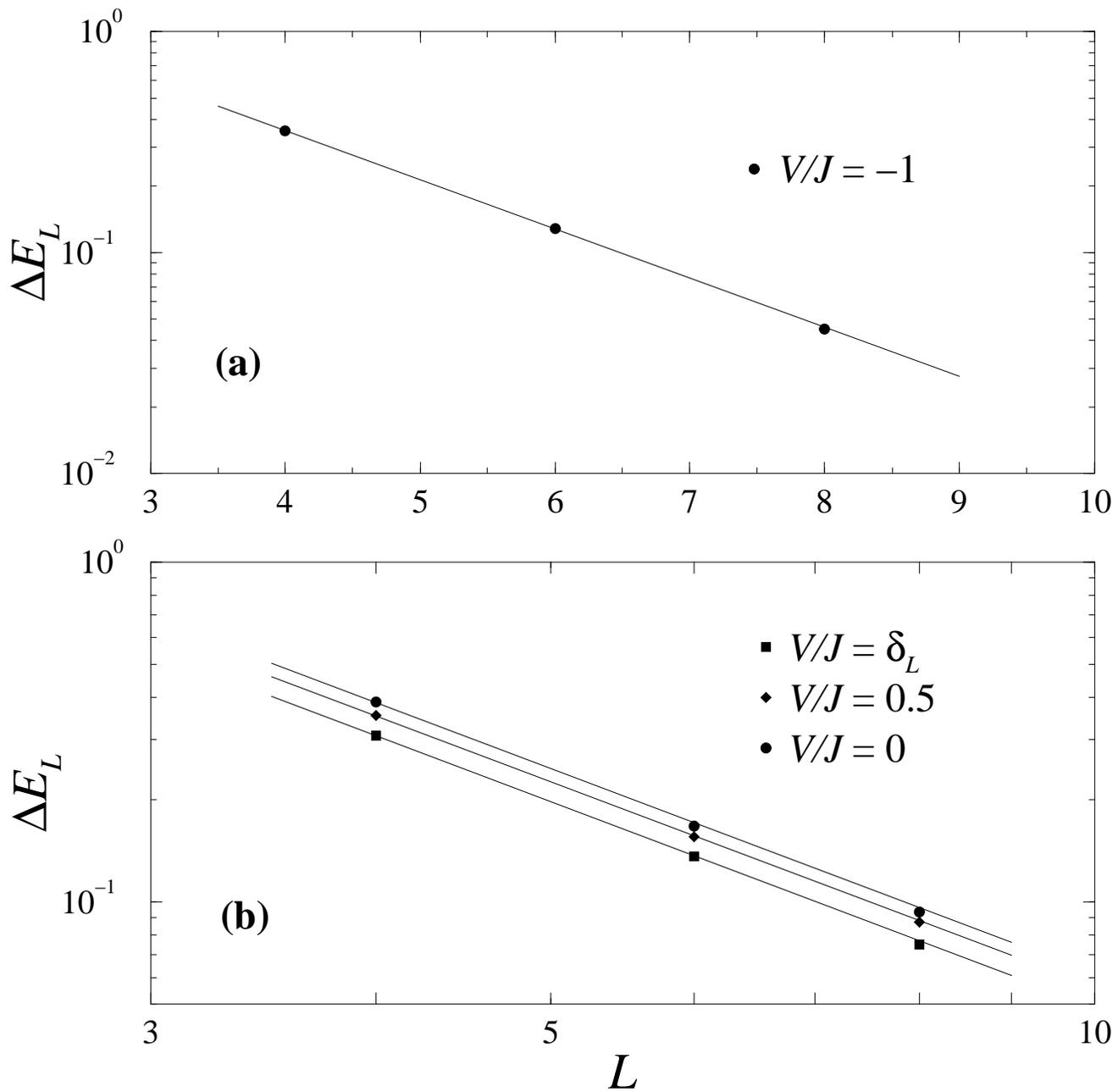
\begin{figure}
\caption{(a) $\Delta E_L$ vs $L$ in semi-logarithmic scale.  
The straight line is the best fit to the data, 
$\Delta E_L=2.742\,e^{-0.511L}$.
(b) Similar plot but in logarithmic scale.  $\delta_L$ is close
to 1 (see Fig.~\protect\ref{deltaL}).  The straight lines
are best fits to the data of the form $\Delta E_L\propto 1/L^2$.}
\label{DeltaE}
\end{figure}

Next we study the energy gap $\Delta E_L$,
which is the energy difference between $E_{(\pi,0)}$ and $E_0$.
Fig.~\ref{DeltaE} shows different plots of $\Delta E_L$
vs $L$ at different $V/J$.
When $V/J$ is close to 1, complication arises because of the spurious
crossing over of the states $E_{(\pi,0)}$ and $E_{(\pi,\pi)}$ at $\delta_L$.
But we find that $\Delta E_L$ is not very sensitive to $V/J$ 
in the region close to $\delta_L$ (see Fig.~\ref{level}).
Consequently we plot $\Delta E_L$ at $V/J=\delta_L$ (instead of
at the same $V/J$ for different $L$) when $V/J$ is close to 1.
From Fig.~\ref{DeltaE} we can clearly distinguish two different behaviors.
At $V/J=-1$, $\Delta E_L$ decays exponentially with $L$.
This is a clear signal for the existence of LRO.
When $0\le V/J<1$, we find that $\Delta E_L$ vanishes with a power law
in $L$.  Best fit to the data shows that the power is approximately
2 in the whole range.  
For $-1<V/J<0$, we are not able to determine whether $\Delta E_L$
has exponential or power law dependence on $L$.
This is probably because of the more serious finite-size effect
when $V/J$ is close to a transition point.
Hence we infer that there are two phases in the range $-1\le V/J<1$.
At some negative $V/J$, a
transition occurs and the dependence of $\Delta E_L$ on $L$
changes from exponential to power law.

\section{Columnar order parameter}
\label{sec:op}

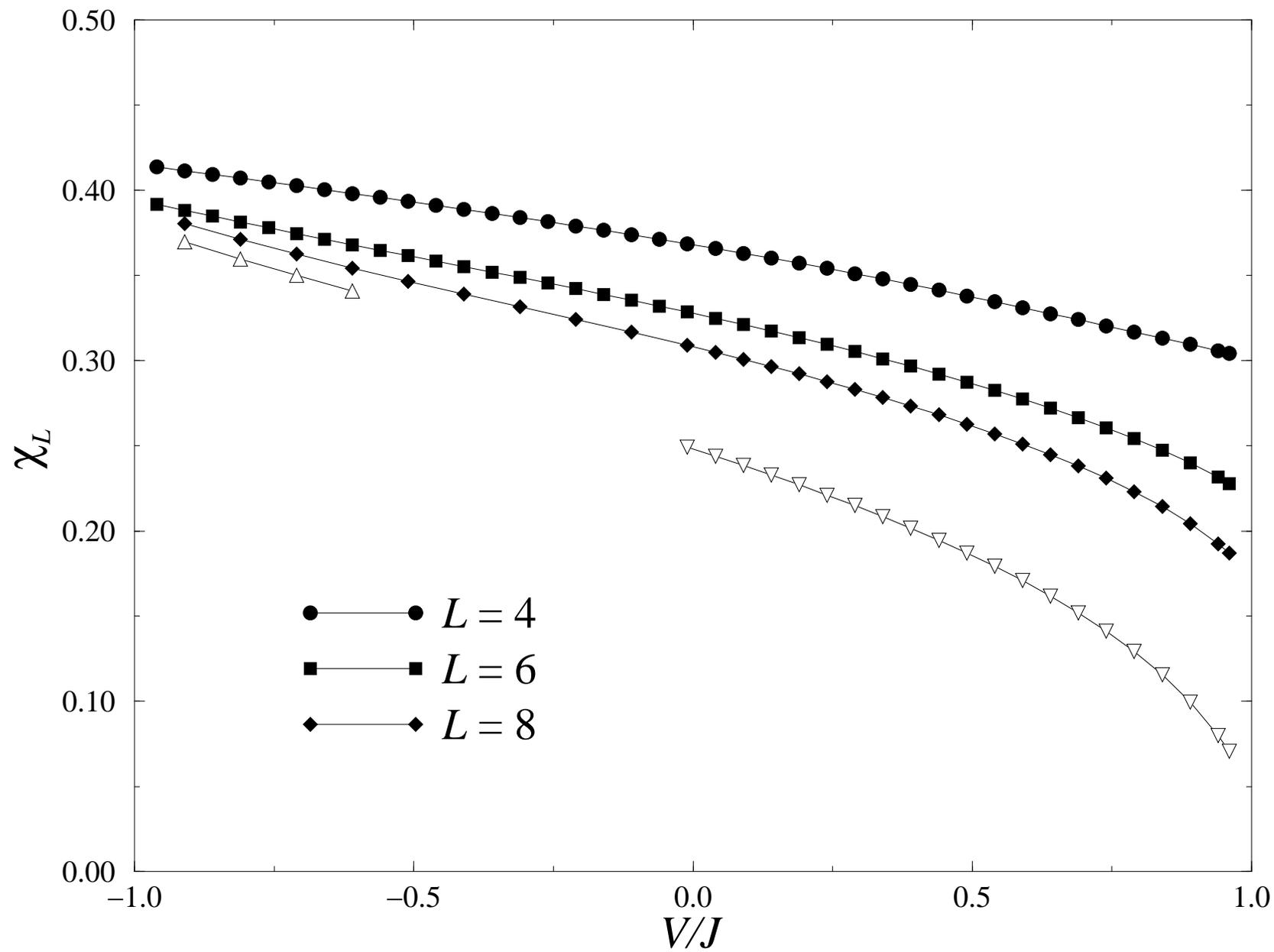
\begin{figure}
\caption{$\chi_L$ vs $V/J$ for $L=4$, 6, 8.  
$\bigtriangleup$ and $\bigtriangledown$ are
$\chi_\infty$ obtained
by extrapolating $\chi_L$ linearly in $1/L^2$ and $1/L$ respectively.}
\label{chivsV}
\end{figure}

The columnar dimer order parameter is defined as\cite{s89}
\begin{eqnarray}
\Psi_{col}({\bf r})&=&(-1)^{r_x}\,\biggl[n({\bf r}+\frac{\hat{\bf x}}{2}) -
n({\bf r}-\frac{\hat{\bf x}}{2})\biggr]\nonumber\\
&+& i\ (-1)^{r_y}\,\biggl[n({\bf r}+\frac{\hat{\bf y}}{2}) -
n({\bf r}-\frac{\hat{\bf y}}{2})\biggr],
\label{psi}
\end{eqnarray}
where $\hat{\bf x}$ and $\hat{\bf y}$ are unit vectors.
The dimer number operator
$n({\bf r}+\frac{\hat{\bf e}}{2})$ is 1 if the site at {\bf r}
and its nearest neighbor at ${\bf r}+\hat{\bf e}$ form a dimer, and
zero otherwise.
In the finite-size study, one defines the order parameter\cite{s89}
\begin{equation}
\chi^2_L=\biggl\langle\biggl|\frac{1}{L^2}\sum_{{\bf r}\in A}\,
\Psi_{col}({\bf r})\biggr|^2\biggr\rangle.
\label{chi}
\end{equation}
For a dimer liquid which has no long-range dimer order,
$\chi_L$ is zero in the large $L$ limit.
If long-range columnar order exists,
$\chi_L$ remains finite at large $L$.
In Fig.~\ref{chivsV} we plot $\chi_L$ versus $V/J$.
When $V/J$ is close to $-1$, $\chi_L$ is approximately linear
in $1/L^2$.  But when $V/J$ is increased, a linear
relation in $1/L$ fits the data better.
Unfortunately we are not
able to determine the point where these two behaviors change
from one to another.
Using these linear relations, we extrapolate
$\chi_L$  to obtain $\chi_\infty$, which
is also plotted in Fig.~\ref{chivsV}.
It shows that $\chi_\infty$ is finite at $V/J<1$, which signifies
the existence of columnar order.
$\chi_\infty$ decreases as $V/J$ approaches 1,
showing that the columnar order is weakening.
But there is no evidence that $\chi_\infty$ is zero at any $V/J<1$.
$\chi_\infty$ clearly shows that the columnar order
persists for $V/J<1$.

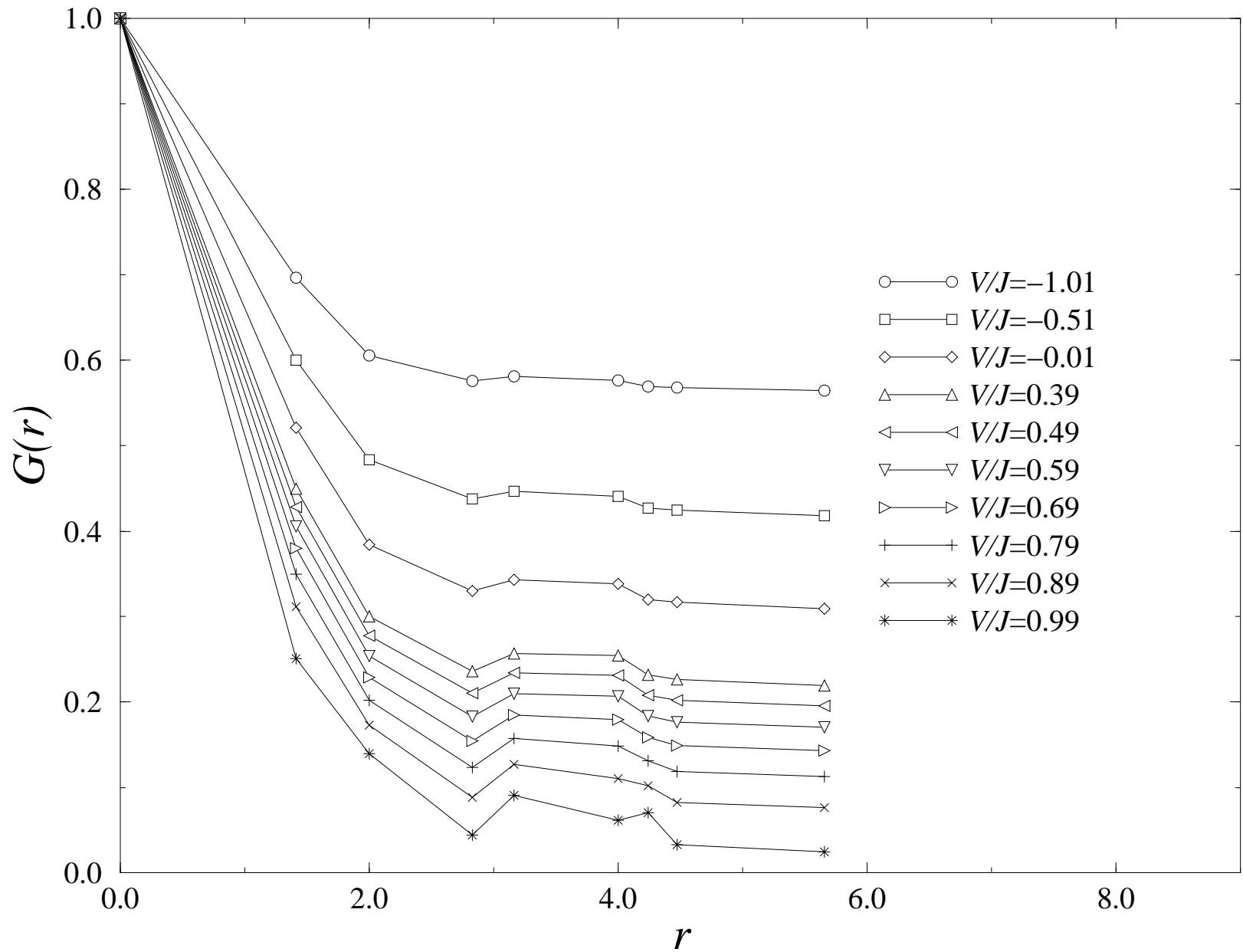
\begin{figure}
\caption{$G(r)$ at different $V/J$ for $L=8$.}
\label{Gr}
\end{figure}

Different dependence of $\chi_L$ on $L$
($\propto 1/L^2$ and $1/L$) may imply  different
orders in the respective regions.  
One defines the correlation function of $\Psi_{col}$,\cite{s89}
\begin{equation}
\label{gofr}
G({\bf r})=\langle\Psi_{col}^*({\bf r}_1)\,
\Psi_{col}({\bf r}_1+{\bf r})\rangle,
\end{equation}
where {\bf r}, {\bf r}$_1$ are in the $A$ sublattice.
In the two regions where $\chi_L$ has finite-size corrections 
$1/L^2$ and $1/L$, $G(r)$ probably
decays exponentially and as a power law in $r$, respectively, to a
non-zero constant value.
Fig.~\ref{Gr} shows $G(r)$ in the $L=8$ system.
Unfortunately, direct observation of how $G(r)$ decays with $r$
is not possible in this system size.
Fig.~\ref{Gr} shows that when $V/J$ is away from 1,
$G(r)$ is
a nonzero constant at large $r$.  
When $V/J$ is close to 1, it is not possible to tell from
Fig.~\ref{Gr} whether $G(r)$ is small or zero at large $r$.
Nevertheless, it tells us that in the range $-1\le V/J< 1$
the long-range order, if exists, should be columnar.

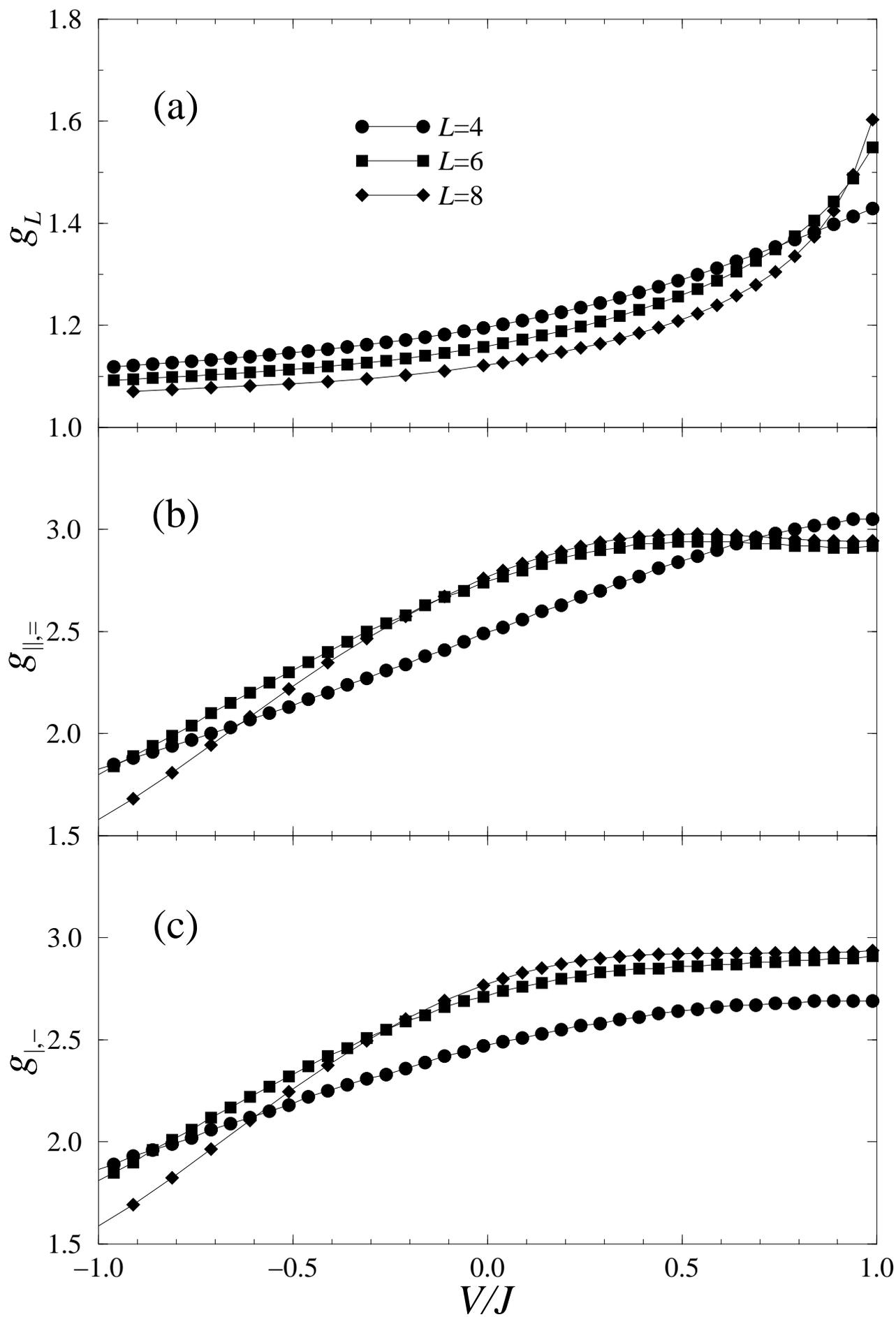
\begin{figure}
\caption{Various cumulants vs $V/J$ for $L=4$, 6, and 8.
(a) $g_L$, (b) $g_{\parallel,=}$, and (c) $g_{|,-}$.}
\label{gvsV}
\end{figure}

As suggested in Ref.~\onlinecite{s89}, a further probe of the columnar
dimer order is provided by the cumulant of the
columnar order parameter,
\begin{equation}
g_L
=\biggl\langle\biggl|\sum_{\bf r}\Psi_{col}({\bf r})\biggr|^4\biggr\rangle \biggl/
\biggl\langle\biggl|\sum_{\bf r}\Psi_{col}({\bf r})\biggr|^2\biggr\rangle^2.
\end{equation}
If long-range columnar dimer order exists, $g_L\rightarrow1$ as
$L\rightarrow\infty$.
We plot $g_L$ versus $V/J$ in Fig.~\ref{gvsV}(a).
Standard finite-size scaling theory\cite{fsc} shows that at phase transition,
$g_L$ at different $L$  cross at a unique point.
Since  $g_L$ do not cross except when $V/J$ is close to 1, Fig.~\ref{gvsV}(a)
tells us that the order parameter $\Psi_{col}$ detects
no transition 
except when $V/J$ is close to 1.  This is the transition
to the dimer liquid (FS) state.  Since our system size $L$ is not
large enough to make $g_L$ cross at one point, it is not possible
to determine the transition point from Fig.~\ref{gvsV}(a).
Nevertheless, our previous finite-size
analysis of  $\Delta E_L$ and
 $\chi_L$ 
strongly suggest that some kind
of columnar LRO persists all the way up to $V/J=1$.
We also note that $g_L$ detects no transition at any
$-1\le V/J<0$, in contrary to the results of $\Delta E_L$
and $\chi_L$.

\section{Disorder within the columns}
\label{sec:random}

It is easy to understand why $g_L$ is not able to detect
the transition at $V/J<0$.
Note that in the plaquette RVB state, $\chi_L$ is
also non-zero in the large $L$ limit.  It can be considered as a  state which
possesses long-range ``columnar order''
(as measured by $\chi_L$), but has disorder within
the established columns.
To detect this disorder
 we introduce an order parameter $M_{\parallel,=}$,
\begin{equation}
M_{\parallel,=} = \frac{1}{L^2}\sum_{\rm plaquettes}\biggl[
n(|\parallel\,\rangle) - n(|=\rangle)\biggr],
\end{equation}
where $n(|\parallel\,\rangle)$ and  $n(|=\rangle)$ are
number operators of vertical 
($\parallel$)
and horizontal ($=$) dimer pairs respectively.
In the perfect columnar state,
$|M_{\parallel,=}|=1/2$, while in the plaquette RVB, dimer liquid and the
staggered state, $M_{\parallel,=}=0$.
Analogous to $\Psi_{col}$, we define the cumulant
\begin{equation}
g_{\parallel,=} = \frac{\langle|M_{\parallel,=}|^4\rangle}
{\langle|M_{\parallel,=}|^2\rangle^2}.
\end{equation}
Again, finite-size scaling theory shows that at the phase transition, 
$g_{\parallel,=}$
at different $L$ should intersect at a unique point.
In Fig.~\ref{gvsV}(b) we plot $g_{\parallel,=}$ vs $V/J$ for
different $L$.  Although the three curves do not intersect
at a single point, the crossing of the curves at $L=6$ and 8
indicates that a transition is possible at some negative
value of $V/J$.  This is consistent with the results
of $\Delta E_L$ and $\chi_L$.  We are not
able to obtain a good estimation for the transition point $V_c/J$.
But very roughly, $V_c/J \sim -0.2$.

We also calculate the moments of $M_{|,-}$,
\begin{equation}
M_{|,-} = \frac{1}{L^2}\sum_{\rm plaquettes}\biggl[n(\,|\,) - n(-)\biggr],
\end{equation}
{\it i.e.} the difference between the number of vertical and
horizontal dimers.  The corresponding cumulant $g_{|,-}$
is shown in Fig.~\ref{gvsV}(c).
We find a similar crossing of $g_{|,-}$ for $L=6,8$
also near $V/J \sim -0.2$. Both $M_{|,-}$ and $M_{\parallel,=}$ should 
be probes of the same ordered columnar to plaquette RVB transition.
We note that for $V>V_c$ both $g_{\parallel,=}$ and $g_{|,-}$
are close to $3$. This value is the expected gaussian moment ratio
for disordered states. This strongly suggests there is no $M_{\parallel,=}$
or $M_{|,-}$ LRO for the whole range of $V>V_c$. For $V/J < -1$
the cumulants approach $1$ for the ordered columnar state. 
In the $L \to \infty$ limit the cumulants should become $1$ for all $V<V_c$.
Fig.~\ref{gvsV}(b) and (c) are consistent with this scenario.

\section{Dimer correlations}
\label{sec:dcorr}

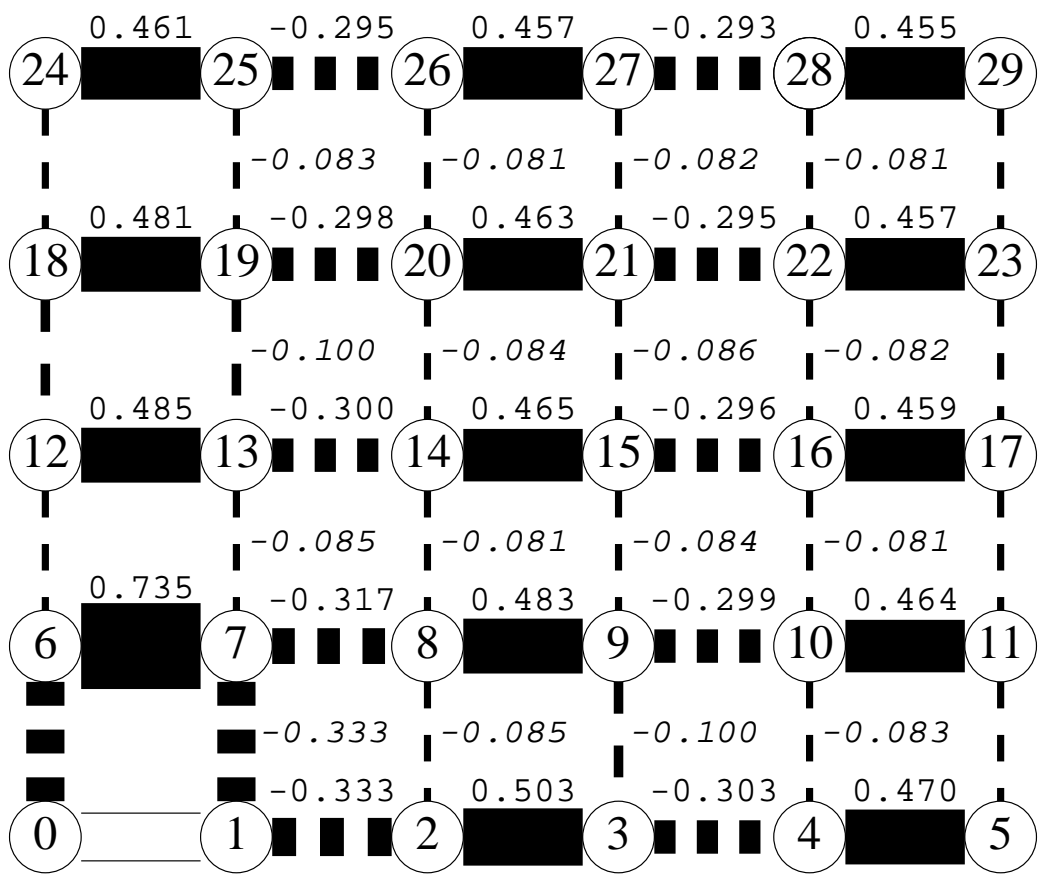
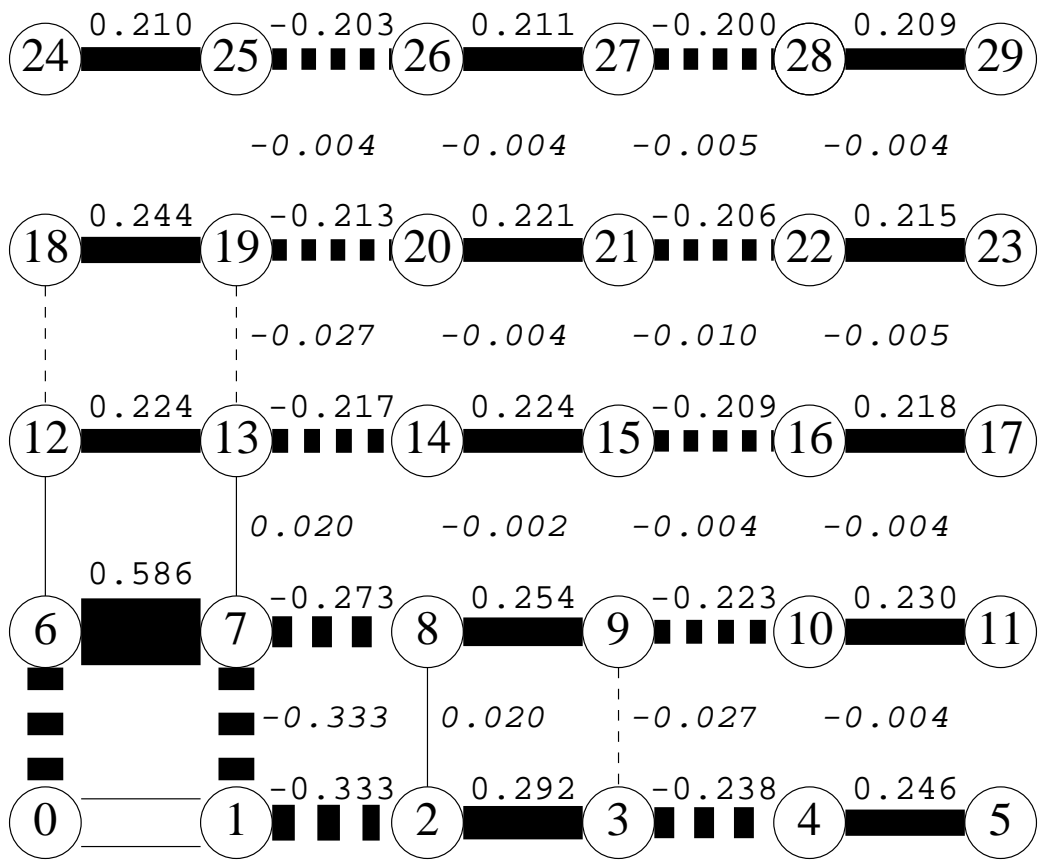
\begin{figure}
\caption{Dimer-dimer correlation $C_{(ij)(kl)}$ in
the $L=8$ system at (a) $V/J=-1$ and  (b) $V/J=0$.
Only a quadrant of the system containing all the inequivalent
dimer pairs is shown.
The reference bond $(ij)$ is represented by a double line.
$C_{(ij)(kl)}$ is proportional to the thickness of the line
joining the pair of sites $k$ and $l$.
Solid line means $C_{(ij)(kl)}>0$, and broken line
means $C_{(ij)(kl)}<0$.
Note that the line widths are in the same scale in both cases.
The values of $C_{(ij)(kl)}$ are shown next to the bonds, where
different fonts are used for vertical and horizontal bonds.}
\label{nij_fig}
\end{figure}

$g_{\parallel,=}$ and $g_{|,-}$ show that the possible transition at $V_c/J$
is between two states with long-range columnar order,
but one has disorder within the columns.
Next we are going to study the dimer correlations in these
two states.
The spatial correlation of the dimers can be displayed by
evaluating the dimer-dimer correlation function,\cite{le93}
\begin{equation}
C_{(ij)(kl)} = \frac{\langle n_{ij}\,n_{kl}\rangle - 
  \langle n_{ij}\rangle^2}
{\langle n^2_{ij}\rangle - \langle n_{ij}\rangle^2},
\label{nij}
\end{equation}
where $n_{ij}\equiv n(\frac{1}{2}({\bf r}_i + {\bf r}_j))$.
If the dimers do not have LRO, $C_{(ij)(kl)}$
should show some short-range correlations, and fall off rapidly
as the separation between the bonds $(ij)$ and $(kl)$ increases.
On the other hand, if the dimers have LRO,
$C_{(ij)(kl)}$ should reflect the pattern of the long-range correlations.
Fig.~\ref{nij_fig} shows the dimer-dimer correlation in the
$L=8$ system at $V/J=-1$ and  0.  
The reference bond $(ij)$ is represented by a double line.
$C_{(ij)(kl)}$ is proportional to the thickness of the line
joining the pair of sites $k$ and $l$.
Solid line means $C_{(ij)(kl)}>0$ (correlation), and broken line
means $C_{(ij)(kl)}<0$ (anti-correlation).
It is obvious that in both cases the overall order is columnar --- those bonds
with positive $C_{(ij)(kl)}$ are arranged in well-defined columns.
This is consistent with the study using $\Psi_{col}({\bf r})$
as the order parameter (Sec.~\ref{sec:op}).
Except for the trivial short-range correlations,
$C_{(ij)(kl)}$ does not fall off significantly with distance.
This is an indication that the correlation is long-range.
A major difference between Fig.~\ref{nij_fig}(a) and (b) is
that in the latter, $C_{(ij)(kl)}$ for vertical $(kl)$
are very close to zero, except for those in contact with $(ij)$.
In the perfect columnar state as shown in Fig.~\ref{dimer}(a),
$C_{(ij)(kl)}=1$ or $-1/3$ when the bond $(kl)$ forms and does
not form dimer respectively. (Here the reference bond $(ij)$ has
been chosen to be one of the horizontal
bonds that form dimer in Fig.~\ref{dimer}(a).)
This is to be compared to the result at 
$V/J=-1$ (Fig.~\ref{nij_fig}(a)).  
In the perfect plaquette RVB state as shown in Fig.~\ref{dimer}(c),
except for those $(kl)$ which are in the same plaquette as $(ij)$,
$C_{(ij)(kl)}=0$ for vertical $(kl)$, and $C_{(ij)(kl)}=\pm1/3$
for horizontal $(kl)$, depending on whether $(kl)$ belongs
to any one of the plaquettes shown in  Fig.~\ref{dimer}(c).
In the same plaquette as $(ij)$, horizontal
and vertical $(kl)$ have $C_{(ij)(kl)}=2/3$ and $-1/3$
respectively.
This is to be compared to the result at 
$V/J=0$ (Fig.~\ref{nij_fig}(b)).  
Hence the dimer correlations strongly suggest that at $V/J=-1$,
the ground state is ordered columns, whereas at $V/J=0$, it is
the plaquette RVB state.

\section{Plaquette RVB state and the transition at $V_{\lowercase{c}}/J$}
\label{sec:nature}

From the above results and discussions, there is evidence that
at $V_c/J$, the ground state of the QDM changes from the ordered columnar
to the plaquette RVB state.
Since this transition does not appear to have been discussed
in the literature, we will discuss it in a little more detail.
One can motivate the existence of this transition
by a simple argument where the QDM is approximately
mapped onto a two-dimensional Ising model in a transverse field.  The $T=0$
transition from ferromagnet to paramagnet in that model corresponds to 
the $V_c/J < 0$ transition of the QDM discussed in the previous sections.
We now outline the steps and implications of this approximate treatment.

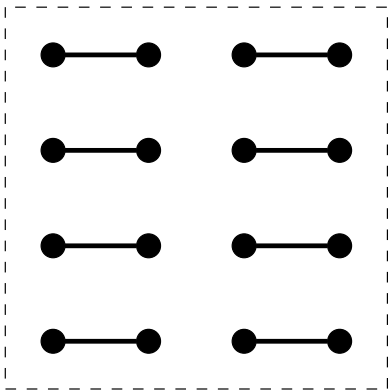
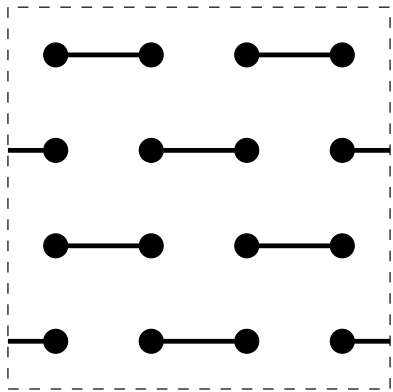
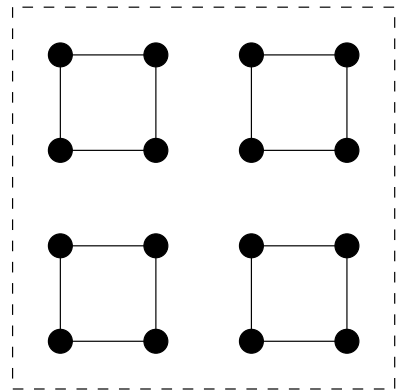
\begin{figure}
\caption{The (a) columnar, (b) staggered, and (c)
plaquette RVB states.}
\label{dimer}
\end{figure}

First consider the case $V/J \ll -1$ where the system tends to the
ordered columnar state in Fig.~\ref{dimer}(a). The basic quantum fluctuations
in the
ground state are induced by the kinetic energy term in (\ref{hamiltonian}),
namely by flipping $|=\rangle$ to $|\parallel\,\rangle$. 
For each dimer populated column
in Fig.~\ref{dimer}(a)
consider {\it every other} plaquette (as in Fig.~\ref{dimer}(c))
to have an index ``$i$" 
and state variable
$\sigma_i^z = 1$ if it is in the state $|=\rangle$ 
and $\sigma_i^z = -1$ if it is 
$|\parallel\,\rangle$. 
Fig.~\ref{dimer}(a) has all $\sigma_i^z = 1$, {\it i.e.} a
ferromagnetic arrangement, whereas Fig.~\ref{dimer}(c)
has all $\sigma_i^z =\pm1$ with probability 1/2, {\it i.e.} a
paramagnetic arrangement.
Note also that the other ferromagnetic Ising configuration, with
all $\sigma_i^z = -1$, gives another perfectly ordered columnar state. 
The collection of states generated by all possible
$\{\sigma_i^z\}$ is, of course, a small subset of the set of all 
close-packed dimer 
configurations.
However, they at least describe the plaquette RVB states,
%two of 
the ordered columnar states, and one half of
the basic flipping excitations. 
Given that our earlier numerical results suggest
columnar order for all $V/J < 1$, we do not worry too much that the 
$\{\sigma_i^z\}$ subset of dimer states are,
by construction, {\it constrained} to have
long-range columnar order,
since we are concerned with the disordering of other degrees of freedom
({\it e.g.} dimer pair arrangements and correlations within
each established column).

Limiting the system to the collection of all possible $\{\sigma_i^z\}$ values,
one can show that the Hamiltonian of the QDM
is, to within an additive constant,
\begin{equation}
{\cal H} = -J \sum_i {\sigma_i^x} + { V \over 4} \sum_{\langle ij \rangle}
\sigma_i^z \sigma_j^z,
\label{approx_hamiltonian}
\end{equation}
where the sums are over every other plaquette as 
shown in Fig.~\ref{dimer}(c). 
$\sigma_i^x$ and $\sigma_i^z$ are Pauli operators.
The kinetic energy term with $J$ now induces flipping of the $\sigma^z$
variables (like a transverse Ising field $h_x$), 
and the potential energy term $V$
introduces interactions (like an Ising exchange integral ${\cal J}_z$)
between neighboring plaquettes as shown
in Fig.~\ref{dimer}(c).\cite{staggered}

For $V/J \ll -1$ all the $\sigma_i^z$ are arranged ferromagnetically,
whereas for some $V_c/J < 0$ the system undergoes a continuous phase
transition (in the universality class of the classical 3D Ising model)
to a state with no LRO in the $\sigma_i^z$ variables. 
Indeed, the case $V=0$ is trivially seen to be perfectly disordered
with respect to the $\sigma_i^z$ variables, and is precisely the
state represented schematically in Fig.~\ref{dimer}(c).
For $V_c < V < 0$ there is some short ranged $\sigma^z$ order, but no LRO.
Since our
$(=)$ $(\parallel)$ pairs of dimers are neither allowed to break apart nor to
move to neighboring plaquettes, this approximation has 
long-range columnar order
(as measured by $\Psi_{col}$) preserved past the transition 
point $V_c$. Our numerical data suggests this to
be the case for the  QDM as well: LRO in $\Psi_{col}$ exists for all
$V/J < 1$ but there is a loss of LRO in $M_{\parallel,=}$
at some negative value of $V$.

One may speculate that the $V/J < 0$ transition in the 
QDM is also in the same universality class of the 3D
Ising model.  Unfortunately the system sizes  we can deal with  are too small
to determine $V_c$ accurately, 
let alone to estimate the critical exponents.

\section{Finite temperature phase diagram}
\label{sec:finiteT}

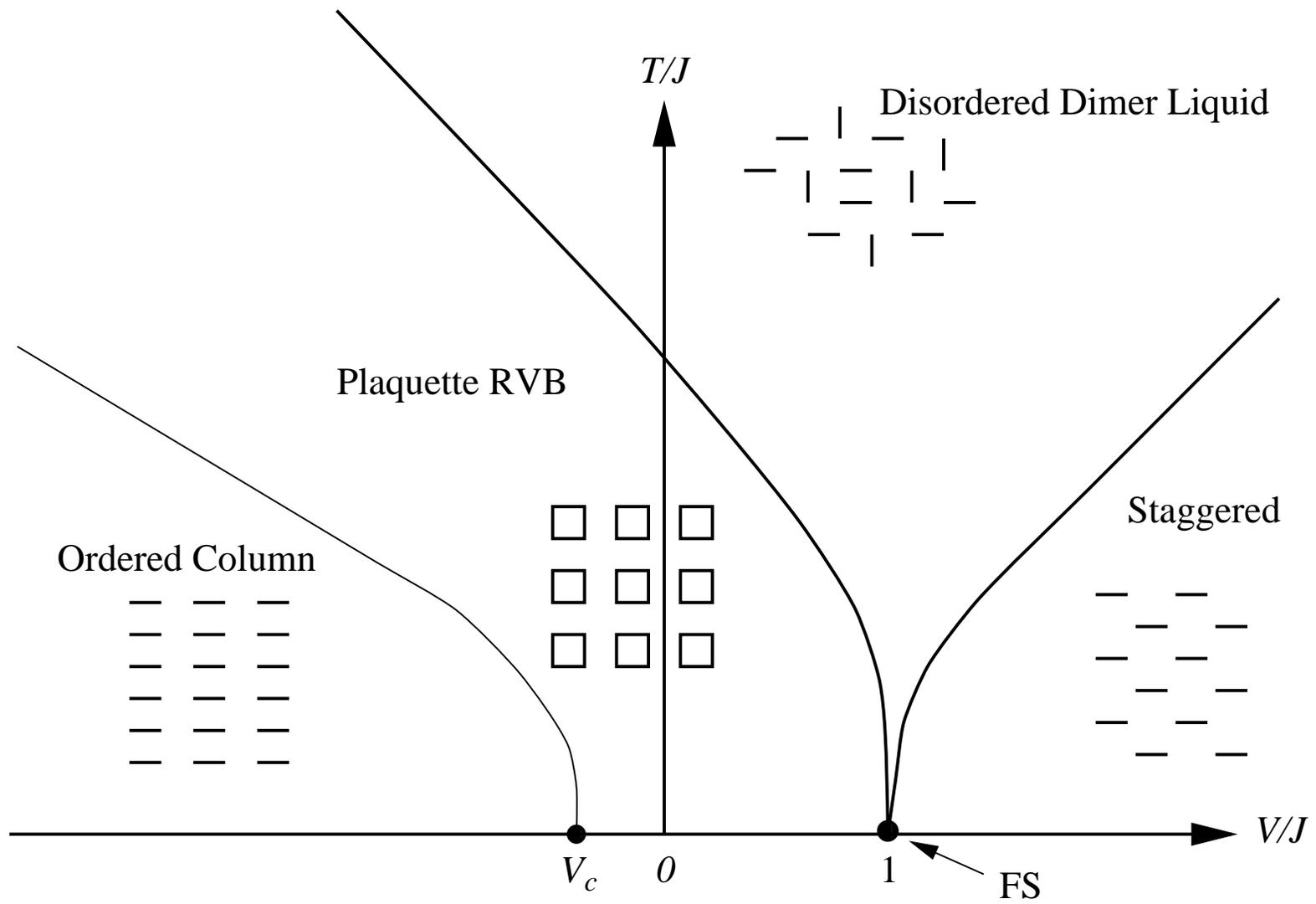
\begin{figure}
\caption{Speculated phase diagram of the QDM.}
\label{pd}
\end{figure}

Although our numerical calculations are limited to the $T=0$ ground state
of the QDM, it is interesting to speculate on the behavior of the 
system at finite temperatures. The simplest, and we feel most likely,
scenario is displayed schematically in Fig.~\ref{pd}.
For any fixed ratio $V/J$, as $T\to \infty$ the only interaction that
matters is the hard-core dimer constraint, and so the FS dimer liquid
will always be obtained at high enough temperature.  We expect that for
fixed $V < V_c$ as $T$ is increased the ordered columnar state will give
way to a plaquette RVB state,\cite{two_plus_one} and at a second higher
temperature the columnar order  will be destroyed as the
system enters the high-temperature dimer liquid phase.
Similarly, for $V>J$ the staggered phase in Fig.~\ref{dimer}(b) will be destroyed
at high enough $T$. The simplest guess is that it goes directly to the
dimer liquid,\cite{first_order} although it is not possible to rule out
other phases.
It seems likely that the high temperature dimer liquid region makes
it all the way down to the $T=0$ axis, but only at the point $V=J$.
It is intriguing %{\bf [SILLY WORD? interesting? amusing?]} 
that 
the $T=\infty$ Fisher-Stephenson (FS) state is {\it exactly} regained at $T=0$ at a single point.

\section{Conclusion}
\label{sec:conclusion}

To conclude, we find no evidence of a dimer liquid state in the
QDM in the range $-1\le V/J<1$.  
In this range, the QDM exists in two states.
At $-1\le V/J<V_c/J$, the ground state
possesses long-range columnar dimer order as suggested
in Fig.~\ref{dimer}(a).  
Its columnar order correlation function
$\langle\Psi_{col}(0)\Psi_{col}({\bf r})\rangle$ is likely to decay to a
constant value exponentially with $r$, and it has
exponentially vanishing energy gap.
This  state should
exist at any $V/J\le -1$, because negative $V$ favors
parallel dimers.
At $V_c/J<V/J<1$, the ground state is the plaquette RVB state
as illustrated in Fig.~\ref{dimer}(c).
Its columnar order correlation
$\langle\Psi_{col}(0)\Psi_{col}({\bf r})\rangle$ is likely to decay to a
constant value as a power law of $r$, and the energy gap
vanishes as a power of $L$.
By approximately mapping the QDM to an Ising model with a transverse
field, we infer that
the transition at $V_c/J$ is  continuous
and possibly in the same universality class as the 3D classical Ising model.
Unfortunately, our system sizes do not allow us to determine
the precise value of $V_c/J$.   Our rough estimate is $V_c/J \sim -0.2$.
We note that we cannot rule out additional transitions in the 
range $V_c/J < V/J < 1$, we can only say the quantities we have 
measured do not indicate any more.

Compared to earlier work\cite{s89} which concluded that a dimer
liquid state {\sl may} exist in $\kappa\le V/J\le 1$ and gave an upper
bound of $0.5$ to $\kappa$, we are able to perform more reliable
finite-size scaling analysis with the addition of the $L=8$ results.
Our results push $\kappa$ to very close to 1 which suggest that
$\kappa$ is in fact 1.  Note that although our results disagree with
Ref.~\onlinecite{o94}, our conclusions have something in common.
Ref.~\onlinecite{o94} claims that the QDM is a dimer liquid
at $V/J=0$, and exhibits a transition from the columnar
to the dimer liquid state at some negative $V/J$.
We also find a transition at some negative $V/J$.\cite{Vc_maybe_zero}
However, this
transition is not from the columnar to dimer liquid state,
but rather to an intermediate state which
we identify to be the plaquette RVB state.

\acknowledgments
PWL thanks C. Zeng for very useful discussions, 
and C. L. Henley  for communicating his results before publication.
This work was supported in part by Hong Kong RGC grant HKUST619/95P.

\appendix
\section{Winding numbers}
\label{app:wind}

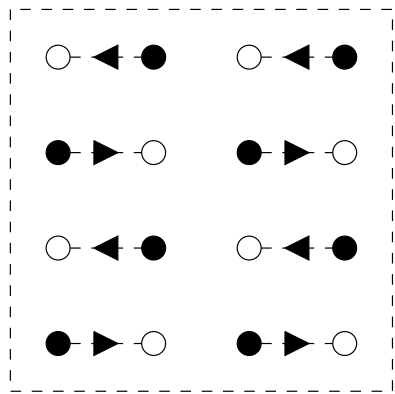
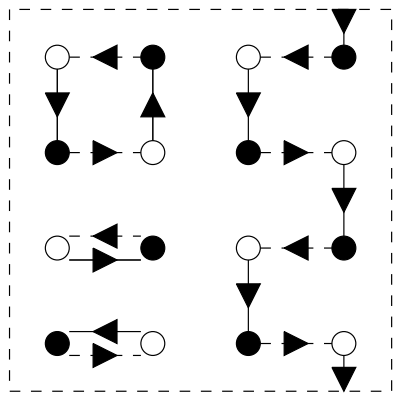
\begin{figure}
\caption{(a) The reference columnar dimer configuration, with arrows
pointing along the dimers from the $A$ sublattice ($\bullet$) to
the $B$ sublattice
($\circ$).  (b) A dimer configuration (indicated by solid lines)
superimposed on (a), but with arrows pointing in the opposite
direction, i.e., from the $B$ sublattice ($\circ$) to 
the $A$ sublattice ($\bullet$).
This dimer configuration has winding numbers $(0,1)$ or $(0,-1)$.}
\label{wind_fig}
\end{figure}

Following Ref.~\onlinecite{rk88} we
use the columnar state as the reference configuration
to define the winding numbers.
In Fig.~\ref{wind_fig}(a), the dimers of such a reference configuration
are indicated by dash lines.  The arrows along the dimers
always point from the $A$ sublattice (solid circles) to the $B$ sublattice
(open circles).  In Fig.~\ref{wind_fig}(b), a dimer configuration
(indicated by solid lines) is superimposed on the reference
configuration, but the arrows along the dimers (solid lines)
point from the $B$ sublattice to the $A$ sublattice.  
This results in directed close paths which may wrap 
around the boundaries
(with periodic boundary conditions).
The winding numbers $\Omega_x$ and $\Omega_y$ are the net number
of loops (clockwise minus counterclockwise) wrapping around the
boundaries in the $x$ and $y$ directions respectively.
Each dimer configuration has a unique pair of winding numbers.
By repeatedly flipping parallel dimer pairs (i.e.,
applying ${\cal H}$) of a dimer configuration, one can generate
all dimer configurations with the same winding numbers.
But no two configurations with different winding numbers
are related by such dimer flipping operations.
Therefore, the Hilbert space spanned by all close-packed dimers
can be divided into subspaces labelled by winding numbers.

To calculate the winding numbers of a given dimer configuration,
we start from a site and traverse the lattice
following the arrows as described above
until we return to the starting site.  We mark all the sites along this
path and count the net number of times it crosses 
the boundaries in the $x$ and $y$ directions.
This procedure is repeated starting from an unmarked site
until all sites are marked. 
It is easy to see that the columnar dimer state in Fig.~\ref{dimer}(a)
has winding numbers $(0,0)$, and the staggered state in Fig.~\ref{dimer}(b)
has winding numbers $(\pm L/2,0)$.

\section{Enumerating dimer configurations}
\label{app:count}

In this Appendix we describe the way we generate all the close-packed
dimer configurations on an $L\times L$ lattice.  Again we divide the
square lattice into two sublattices $A$ and $B$.
At close-pack, each site is one end of exactly one
dimer.  Consequently, a dimer configuration can be represented by
specifying the direction ($\pm\hat{\bf x}$ or $\pm\hat{\bf y}$) of the
dimer attached to each site in the $A$ sublattice.  There are
$2^{L^2}$ such combinations.  But most of them are forbidden
because some sites in the $B$ sublattice
have zero or more than one dimer attached.
The most straightforward way to enumerate all the dimer configurations
is to scan through these $2^{L^2}$ combinations to identify
the allowed configurations.  But this takes prohibitively long for
$L=8$.  We need a clever way to identify and 
eliminate the forbidden configurations
quickly.

We further divide the $A$ sublattice into two quarter-lattices
$A_0$ and $A_1$.  
Each site takes up two bits in a computer word,
with the four binary values 00, 01, 10, and 11 representing
the direction of the dimer attached to it.
We divide our task into two steps.  First we generate allowed
configurations for $A_0$ alone, without considering $A_1$.
Then we generate the allowed configurations for $A_1$ under
the constraints of the $A_0$ configurations.
The $A_0$ and $A_1$ configurations together give all the
allowed dimer configurations.

$A_0$ has $L^2/4$ sites.  Each $A_0$ configuration is represented 
in the computer
by an integer word with at least $2\times(L^2/4)=L^2/2$ bits.  
The $2j$th and $2j+1$th bits of the integer word represent the direction
of the dimer attached to the $j$th site of $A_0$.
In this way each $A_0$ configuration is uniquely associated with
an integer value.
To scan through all the possible $A_0$ configurations, 
we start from an integer value $2^{L^2/2}-1$ whose binary representation
has 1s in the lower $L^2/2$ bits, and 0s otherwise.
This integer value corresponds to the $A_0$ configuration with all
dimers pointing in the same direction.  Successively decreasing this integer
value by 1 is equivalent to rotating the dimers in all possible directions
starting from the 0th, 1st, $\ldots$ sites.
In this way all possible $A_0$ configurations can be generated.
For each integer value (i.e., 
possible $A_0$ configuration), we check for conflict
starting from the last site.
If at some stage we find that the dimers attached to
the $i$th and $j$th sites ($i>j$) of $A_0$ are conflicting (both connect to
the same site in the $B$ sublattice), we can skip all the bits
lower than the $2j$th bit, i.e., instead of decrementing the integer
value by 1, we can go to the next possible configuration
by changing the $2j$th and $2j+1$th bits.
This substantially decreases the run time.  For $L=8$,
$A_0$ has 9\,983\,558 allowed configurations.

The set of allowed configurations for $A_1$ is the same as that for $A_0$.
Naively we can match these two sets of configurations and eliminate
the forbidden ones.  But this involves $(9\,983\,558)^2\sim 10^{14}$ 
steps for $L=8$
and will take prohibitively long.
We must fully utilize the geometrical constraints imposed by the
$A_0$ configurations when enumerating the possible $A_1$ configurations.
For a particular $A_0$ configuration, we mark the allowed dimer
directions for each site in $A_1$ under the constraints
imposed by that $A_0$ configuration.  The total number of combinations
is substantially smaller than $2^{L^2/2}$.
We then scan through all these combinations in the same manner 
as in enumerating the $A_0$ configurations.  
Repeating the same procedure generates
all the allowed dimer configurations.
For $L=4,$ 6, and 8, the number of allowed dimer configurations are
272, 90\,176, and 311\,853\,312 respectively.  For $L=8$, all
the allowed dimer configurations can be enumerated in about four
hours using an HP 735 workstation.

% now the references. delete or change fake bibitem. delete next three
%   lines and directly read in your .bbl file if you use bibtex.

\begin{references}
\bibitem{staggered} We note that if the ``expansion'' is performed
about staggered columns (every other dimer column shifted by one plaquette)
additional terms proportional to $V\sum \sigma_i^z$ will appear in ${\cal H}$,
but this does not change our conclusions that a disordering transition
occurs for some $V/J < 0$.
\bibitem{two_plus_one} Recalling our simple transverse Ising model
description, this $T>0$ transition to plaquette RVB state should
be in the 2D classical Ising model, not 3D. Even if the transverse Ising
picture breaks down, it should be generally true that the  quantum
phase transition of the QDM at $T=0$, $V=V_c$ is in a 2+1 dimensional classical
system universality class, while the rest of the $T>0$ critical line
is in some 2-dimensional system universality class.
\bibitem{first_order} The substantial difference in symmetry
and winding numbers between the staggered phase and the dimer liquid
may mean that the transition between them is first-order.
\bibitem{Vc_maybe_zero} Indeed, our analysis cannot rule out $V_c=0$.
\end{references}
\end{document}